# The role of type II spicules in the upper solar atmosphere

J. A. Klimchuk[1]

[1] We examine the suggestion that most of the hot plasma in the Sun's corona comes from type II spicule material that is heated as it is ejected from the chromosphere. This contrasts with the traditional view that the corona is filled via chromospheric evaporation that results from coronal heating. We explore the observational consequences of a hypothetical spicule dominated corona and conclude from the large discrepancy between predicted and actual observations that only a small fraction of the hot plasma can be supplied by spicules (<2% in active regions, <5% in the quiet Sun, and <8% in coronal holes). The red-blue asymmetries of EUV spectral lines and the ratio of lower transition region (LTR; $T \leq 0.1$ MK) to coronal emission measures are both predicted to be 2 orders of magnitude larger than observed. Furthermore, hot spicule material would cool dramatically by adiabatic expansion as it rises into the corona, so substantial coronal heating would be needed to maintain the high temperatures that are seen at all altitudes. We suggest that the corona contains a mixture of thin strands, some of which are populated by spicule injections, but most of which are not. A majority of the observed hot emission originates in non-spicule strands and is explained by traditional coronal heating models. However, since these models predict far too little emission from the LTR, most of this emission comes from the bulk of the spicule material that is only weakly heated and visible in He II (304 Å) as it falls back to the surface.

## 1. Introduction

[2] One of the great challenges facing space science has been to explain the million degree plasma observed in the solar corona and the coronae of other late-type stars. The underlying atmosphere is much colder, so the hot plasma cannot be energized by a thermal conduction flux from below. It is generally assumed that some mechanism such as magnetic reconnection or waves heats the plasma locally to high temperatures. This traditional view is now being challenged by new observations. Small finger-liked ejections of chromospheric material have been discovered and given the name type II spicules [*De Pontieu et al.*, 2007]. They differ from classical spicules in that they are thinner (<200 km), faster (50–150 km s$^{-1}$), and shorter-lived (10–150 s). Another important property is that a fraction of the cold mass is heated to coronal temperatures as it is ejected [*De Pontieu et al.*, 2011] (although see *Madjarska et al.* [2011] and *Vanninathan et al.* [2012]). This raises the possibility that much or even most of the plasma observed in the corona comes from type II spicules and that heating in the corona itself is unnecessary [*De Pontieu et al.*, 2009, 2011].

[3] The proposal that spicules supply the corona with its mass is not new [*Athay and Holzer*, 1982], but the idea was largely rejected because classical spicules are not observed to reach coronal temperatures. Although it is not universally accepted that type II spicules are different from classical spicules [*Zhang et al.*, 2012; *Sterling et al.*, 2010], the new observations show that at least some hot material is transported to the corona during some ejections.

[4] The primary purpose of this paper is to make an initial assessment of whether type II spicules can explain the corona [*Klimchuk*, 2011]. Do they supply the corona with a majority of its hot plasma or is coronal heating and the associated evaporation of chromospheric material still the dominant process? We henceforth reserve the term "coronal heating" explicitly for energy deposition that takes place above the chromosphere (i.e., in plasmas that begin with temperatures >10$^4$ K). The heating that raises the temperature of cold spicule material to coronal values is not included in this definition.

[5] Observations show that only a small fraction of the spicule mass is heated to high temperatures, usually at the tip. The rest falls back to the surface in a much cooler state. Our analysis does not concern the origin of spicules or the cause of the heating, but rather the evolution of the hot plasma after it is created. Our approach is to assume that all coronal plasma comes from spicules, with no coronal heating, and to

[1]Heliophysics Division, NASA Goddard Space Flight Center, Greenbelt, Maryland, USA.

Corresponding author: J. A. Klimchuk, Heliophysics Division, NASA Goddard Space Flight Center, 8800 Greenbelt Rd., Greenbelt, MD 20771, USA. (james.a.klimchuk@nasa.gov)

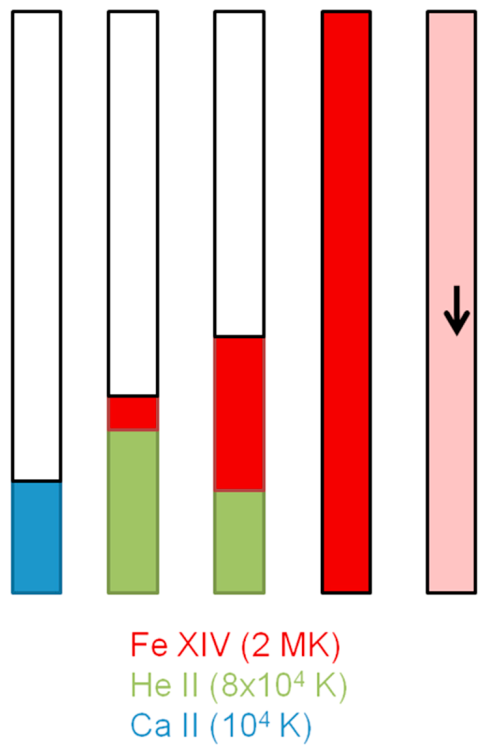

**Figure 1.** Schematic representation of the plasma evolution under type II spicule scenario A (most common). Snapshots are shown, with time increasing from left to right. First snapshot: rising spicule is visible in Ca II (3968 Å) and $H_\alpha$ (blue). Second snapshot: most of the spicule is heated to ≲0.1 MK (green) and becomes visible in He II (304 Å) as it continues to rise; tip is heated to ∼2 MK (red) and becomes visible in Fe XIV (274 Å). Third snapshot: warm material falls and hot material expands into the corona. Fourth snapshot: strand becomes filled with hot material. Fifth snapshot: material cools and drains slowly back to surface.

parts of our analysis resemble the approach of *De Pontieu et al.* [2009].

## 2. Model

[7] As described in *De Pontieu et al.* [2011], most type II spicules are observed to evolve in a manner shown schematically in Figure 1. Each vertical column represents a snapshot in the evolution, with time increasing from left to right. Ca II H (3968 Å) movies from the Solar Optical Telescope (SOT) on Hinode reveal a thin (diameter $d \approx 200$ km) jet of cool material ($T \approx 10^4$ K) extending upward from the limb at an apparent velocity $v \approx 100$ km s$^{-1}$ (blue). The disk counterparts are believed to be rapid blueshift events (RBEs) seen in $H_\alpha$ [*Rouppe van der Voort et al.*, 2009]. Most of the material then disappears in Ca II and $H_\alpha$ as it is heated to approximately $8 \times 10^4$ K and becomes visible in He II (304 Å), as observed by the Atmospheric Imaging Assembly (AIA) on the Solar Dynamics Observatory (SDO) (green). Some of the material at the top (the upper fraction $\delta \approx 10\%$ by length) is heated to much higher temperatures of approximately 1–2 MK and is visible in the 171, 193, and 211 channels of AIA, which are dominated by Fe IX, Fe XII, and Fe XIV, respectively (red). The warm He II spicule falls back to the surface after reaching a maximum height $h_s \approx 10{,}000$ km, somewhat higher than that of the cool Ca II spicule from which it is transformed. The hot material continues to rise rapidly upward into the corona and fades from view. (The dark red in the figure indicates temperature, not brightness.) Note that the actual length-to-diameter aspect ratio is a factor of 10 greater than indicated in the figure. Henceforth, we use the shorthand "spicules" to refer to type II spicules.

[8] Although the physical origin of spicules has not yet been determined, it would seem that the evolution after they are formed can be reasonably well described by one-dimensional hydrodynamics. Spicules are highly columnated structures that expand upward along their primary axis. Almost certainly they are aligned with the magnetic field. If the field is untwisted, then the Lorentz force vanishes along the axis, and a non-magnetic driver is implied. Propagating twist could exert an upward magnetic force, but it is more likely that the spicule plasma is ejected along the field by a locally enhanced gas pressure at its base. Such a high pressure region would be produced, for example, when a reconnection outflow jet in the chromosphere decelerates and its kinetic energy is thermalized. It could also be produced by a local squeezing of the flux tube during the interaction with emerging flux [*Martínez-Sykora et al.*, 2011b].

[9] It is important to realize that spicules cannot be produced by an impulsive heating event in the corona or transition region. Such an event drives chromospheric evaporation, but this is entirely different from a spicule. Only hot material rises into the corona during evaporation. There is no cold jet. The differences between spicules and evaporation are discussed in detail in Appendix A.

[10] The primary model that we examine in this paper describes the hydrodynamic field-aligned evolution of the plug of hot (2 MK) plasma that appears abruptly at the top of the spicule (second snapshot in Figure 1). It is presumably much denser and possibly also much hotter than the ambient coronal plasma above, so it should expand rapidly upward into this lower pressure region. The third and fourth snapshots in

consider the observational consequences of this assumption. We will show that predicted observations are in gross disagreement with actual observations, and conclude from this that the *a priori* assumption must be incorrect: spicules provide only a small fraction of the hot plasma that exists in the corona, and coronal heating is mandatory. Our analysis also leads us to suggest that most of the bright emission from the LTR ($T \lesssim 0.1$ MK) comes from the bulk of the spicule material that falls back to the surface after being only weakly heated. Traditional models have difficulty explaining both the brightness of this emission and its rapid redshifts (although see *Hansteen et al.* [2010]).

[6] Although our analytical treatment is highly simplistic, it captures the essential physics of the spicule phenomenon as it is currently understood based on available observations. Our conclusions must nonetheless be judged relative to the simplifying assumptions that we make. Hydrodynamic and MHD simulations as well as new observations will ultimately test the validity of these assumptions. What we offer here is a reasonable and we believe very meaningful early attempt at addressing this important problem. We note that

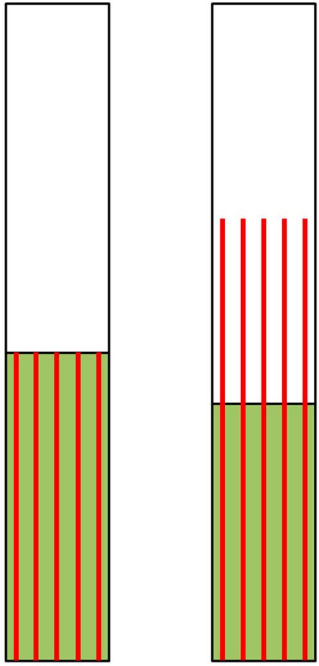

**Figure 2.** Schematic representation of the plasma evolution under type II spicule scenario B (least common). (left) Spatially unresolved warm (green) and hot (red) sub-strands appear together. (right) Warm material falls and hot material expands into the corona.

Figure 1 represents this expansion and eventual filling of the strand. Because we are testing the hypothesis that spicules explain the corona, the temperature must remain near 2 MK as the strand fills. This poses a serious challenge to the hypothesis that there is no coronal heating, since adiabatic cooling is severe. We discuss this later, but for now simply assume that the expanding plasma stays at a temperature near 2 MK. Once the strand is filled, it will subsequently cool by radiation and thermal conduction and drain back down to the chromosphere. This is represented in the fifth snapshot. Eventually the strand will become "empty" to match the initial conditions before the spicule was created.

[11] The sequence of events described above, with hot emission appearing near the top of the warm He II spicule, is most common, and we refer to it as scenario A. A minority of spicules have a slightly different behavior in which the hot emission appears along the full length of the warm structure [*De Pontieu et al.*, 2011; S. W. McIntosh, personal communication, 2011]. For these, we envision the configuration shown in Figure 2. The spicule contains spatially unresolved strands, most of which are near $8 \times 10^4$ K and some of which are near 2 MK. In this case the parameter $\delta \approx 1$ since the hot plasma occupies the full length of the spicule. The expansion of the plasma and subsequent cooling and draining are the same as before. We call this scenario B. Note that there is likely to be unresolved substructure in scenario A as well. The figures are conceptual idealizations.

## 2.1. Blue Wing to Line Core Intensity Ratio

[12] If a substantial amount of hot spicule material flows into the corona at speeds of approximately 100 km s$^{-1}$, then we should see evidence for it in the blue wings of spectral lines such as Fe XIV (274 Å) when observed on the disk. Such evidence has been reported as red-blue (RB) asymmetries in the line profile by *Hara et al.* [2008], *De Pontieu et al.* [2009], *McIntosh and De Pontieu* [2009], *De Pontieu et al.* [2011], and *Tian et al.* [2011]. These studies find that the excess intensity in the blue wing is approximately 5% of the intensity in the line core. *Doschek* [2012] reports intensity ratios of generally <5% in Fe XII (195 Å). Other researchers have had difficulty finding any detectable RB asymmetry [e.g., *Tripathi et al.*, 2012]. The asymmetries seem to be most pronounced at the periphery of active regions.

[13] The blue wing to line core intensity ratio provides an important constraint on the amount of coronal plasma that comes from spicules. Under our hypothesis that all coronal plasma has a spicule origin, we have a simple interpretation for both the wing and core emission. The wing emission is produced by the hot spicule material as it is rapidly expanding upward, and the core emission is produced by the same material as it is cooling and slowly draining (the line core should be weakly redshifted). Of course we do not observe rising and falling material from the same event at the same time. Spicules are very small features, however, and there are likely to be several unresolved events contributing to an observed line profile. Typical EUV spectrometer observations have an effective spatial resolution of 1000 km or larger, depending on pixel summing, and exposure times of several tens of seconds. If spicules occur randomly, then an observed line profile can be approximated by the time-averaged emission from a single event that has gone through a complete upflow and downflow cycle. To predict the blue wing to line core intensity ratio, we must estimate the time-integrated emission separately for the upflow and downflow. We begin with the upflow, and consider observations well away from the limb.

[14] Let $h_0 = \delta h_s$ be the length along the magnetic strand (vertical thickness) of the hot spicule material just after being heated to 2 MK. This heating occurs very rapidly based on the observation that Ca II spicules disappear from view in only 5–20 s [*De Pontieu et al.*, 2007]. Let $n_0$ be the electron number density at this time. Since mass is conserved as the hot material expands,

$$n(t)h(t) = n_0 h_0. \quad (1)$$

For a constant expansion velocity, $v$,

$$h(t) = h_0 + vt. \quad (2)$$

If the temperature remains constant during the expansion, then the intensity of an emission line is proportional to the column emission measure, $n^2 h$. Substituting from above, we have

$$EM(t) = EM_0 \frac{1}{1 + vt/h_0}, \quad (3)$$

where $EM_0 = n_0^2 h_0$ is the initial value. The time-integrated emission measure of the expanding hot plasma column at time $t$ is therefore

$$EM^*(t) = \int_0^t EM(t) dt = EM_0 \frac{h_0}{v} \ln\left(1 + \frac{vt}{h_0}\right). \quad (4)$$

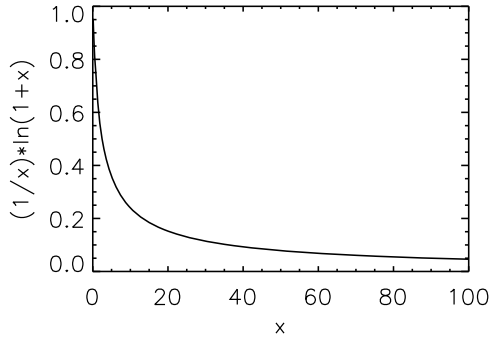

**Figure 3.** The function $f(x) = (1/x)\ln(1 + x)$ used in equation (5).

A typical coronal scale height $h_c = 5 \times 10^4$ km is filled in upflow time $\tau_u = h_c/v = 500$ s. We take this to be the end of the expansion phase, so the time-integrated emission measure of the upflow is

$$EM_u^* = EM^*(\tau_u) = \tau_u EM_0 f(x), \quad (5)$$

where $x = h_c/h_0$ and $f(x) = (1/x)\ln(1 + x)$. As shown in Figure 3, $f(x) \approx 0.1$ for the reasonable range $0.05 \leq \delta \leq 0.3$, corresponding to $17 \leq x \leq 100$. Recall that $\delta = h_0/h_s$ is the fraction of the spicule that is heated to coronal temperatures. The time-integrated emission measure of the upflow is then

$$EM_u^* \approx 0.1 \tau_u n_0^2 \delta h_s. \quad (6)$$

The coefficient in the expression becomes 0.4 for scenario B, where $\delta = 1.0$.

[15] After filling the loop strand, the plasma cools and drains. To explain the corona, its density must equal the observed characteristic density, $n_c$. We can write the emission measure as $n_c^2 h_c A$, where an area factor $A$ is introduced to account for the difference in the average cross-sectional areas of the upflow and downflow. The magnetic field does not change between the upflow and downflow, but the integrated emission from the upflow is dominated by the earliest times, when the material is still low in the corona, whereas the emission from the downflow comes more or less uniformly from the full length of the strand. If the strand expands with height, then $A > 1$. *De Pontieu et al.* [2009, 2011] suggest $A = 3$, which seems reasonable. We note that both the upflow and downflow are well above the throat of rapid magnetic expansion at the base of the strand that is associated with the sharp transition from high to low plasma $\beta$.

[16] The downflowing plasma will be visible in a line like Fe XIV (274 Å) only for a characteristic cooling time, $\tau_{cool}$. This is true whether the cooling begins near 2 MK or at a higher temperature. The time-integrated emission measure appropriate for computing the intensity is therefore

$$EM_d^* = \tau_{cool} n_c^2 h_c A. \quad (7)$$

If we assume cooling by radiation and use the optically thin radiative loss function at 2 MK in *Klimchuk et al.* [2008], we have

$$\tau_{cool} < \tau_{rad} = 6.6 \times 10^{12} n_c^{-1}. \quad (8)$$

The inequality accounts for the fact that we have ignored cooling from thermal conduction and enthalpy, which can be significant [*Bradshaw and Cargill*, 2010]. The time-integrated emission measure of the downflow is finally

$$EM_d^* < 6.6 \times 10^{12} n_c h_c A. \quad (9)$$

[17] The least known quantity in our analysis is the initial density of the hot plasma when it first appears at the tip of the spicule, $n_0$. It can be eliminated from equation (6) using conservation of mass:

$$n_0 \delta h_s = n_c h_c A. \quad (10)$$

A typical coronal density of $n_c = 10^9$ cm$^{-3}$ implies an initial density in the range $5 \times 10^{10} \leq n_0 \leq 3 \times 10^{11}$ cm$^{-3}$ for $0.3 \geq \delta \geq 0.05$. This similar to the $10^{11}$ cm$^{-3}$ density that is measured at chromospheric temperatures in ordinary spicules [*Beckers*, 1972; *Sterling*, 2000]. A direct measurement of $n_0$ using the blue wing emission in density sensitive line pairs would be extremely useful and is something that we are currently pursuing (S. Patsourakos et al., manuscript in preparation, 2012). A major discrepancy with the predicted value above would be strong evidence that a majority of coronal plasma cannot come from spicules.

[18] Combining equations (6), (9), and (10), and using $\tau_u = h_c/v$, we obtain an expression for the ratio of the time-integrated emission measures of the upflow and downflow under scenario A:

$$R = \frac{EM_u^*}{EM_d^*} > 1.5 \times 10^{-14} \frac{n_c h_c^2 A}{\delta h_s v}. \quad (11)$$

The only difference under scenario B is that the coefficient becomes $6 \times 10^{-14}$, since then $\delta = 1$ and $f(x) = 0.36$ in equation (5).

[19] What we really seek is the ratio of blue wing to line core intensities. This ratio will equal the ratio of emission measures only if the upflow and downflow have similar temperature. By our choice of cooling time, we have computed a downflow emission measure appropriate to a 2 MK spectral line. It applies if the cooling/downflow phase begins at this temperature or higher. However, a significantly higher temperature would be inconsistent with the observed emission measure distribution of the corona, which peaks near 2–3 MK. It has been shown that a radiatively cooling plasma has greatest emission measure at its initial temperature if the radiation loss function has a negative slope, as it does in this temperature range [*Sturrock et al.*, 1990; *Bradshaw et al.*, 2012].

[20] Since we are attempting to explain the hot corona in the absence of coronal heating, and since the upflowing plasma will cool as it expands, the initial temperature of the upflow must be at least 2 MK. The cooling effects of expansion are quite dramatic, even without thermal conduction and radiation. For purely adiabatic expansion, $PV^\gamma$ is a constant, where $V$ is the volume of the expanding column and $\gamma = 5/3$ is the ratio of specific heats. It is easy to show that an initial temperature $T_0$ leads to a final temperature

$$T_c = \left(\frac{\delta h_s}{h_c A}\right)^{2/3} T_0. \quad (12)$$

**Table 1.** Blue Wing to Line Core Intensity Ratio, $R$

|  | Scenario A | Scenario B | Observed |
|---|---|---|---|
| Active Region | >3.4 | >1.4 | ≤0.05 |
| Quiet Sun | >1.1 | >0.46 | ≤0.05 |
| Coronal Hole | >0.68 | >0.27 | ≤0.05 |

For $T_0 = 2$ MK, the final temperature is $7 \times 10^4$ K in scenario A and $3 \times 10^5$ K in scenario B. It therefore requires essentially as much energy to maintain the plasma at 2 MK as it does to produce it in the first place. This must come from coronal heating. Converting the initial kinetic energy of the ejection into thermal energy has minimal effect, as discussed in Appendix B.

[21] In order for there to be no coronal heating, the upflow must begin at a much higher temperature than 2 MK, so that it cools to that value by the end of the expansion phase. From equation (12), the required starting temperature would be 58 MK in scenario A and 13 MK in scenario B. This can ruled out, since the ultra hot plasma would be much brighter than observed, as we now show.

[22] Consider a spectral line that is sensitive to the initial temperature $T_0$. The expanding plasma will emit in this line until the temperature drops by approximately a factor of two. From equation (12), the length of the hot column at this time is

$$h = \frac{2^{3/2} \delta h_s}{A_h}, \qquad (13)$$

where $A_h$ is the average cross sectional area over this length compared to the initial length $h_0$. Using equations (4) and (10), we find that the time-integrated emission measure at this stage in the expansion is

$$EM_h^* = \ln\left(\frac{2^{3/2}}{A_h}\right) \frac{(n_c h_c A)^2}{v}. \qquad (14)$$

Combining with equation(9), we get

$$\frac{EM_h^*}{EM_d^*} > 1.5 \times 10^{-13} \ln\left(\frac{2^{3/2}}{A_h}\right) \frac{n_c h_c A}{v} \qquad (15)$$

for the ratio with the 2 MK downflow emission measure. Taking $n_c = 3 \times 10^9$ cm$^{-3}$, $h_c = 5 \times 10^4$ km, $v = 100$ km s$^{-1}$, $A = 3$, and $A_h = 1$ (since $h$–$h_0$ is only a few thousand kilometers), we have $EM_h^* > 0.7 EM_d^*$. Thus, the emission measure at initial temperature $T_0$ is comparable to the emission measure at 2 MK. Since the actual emission measure distribution of the corona peaks near 2–3 MK and decreases steeply at higher temperatures [*Patsourakos and Klimchuk*, 2009; *Reale et al.*, 2009; *Tripathi et al.*, 2011; *Winebarger et al.*, 2011; *Warren et al.*, 2012; *Schmelz and Pathak*, 2012], we conclude that $T_0 \approx 2$ MK. The expanding plasma cannot begin at a very high temperature. Coronal heating is necessary to compensate for the severe cooling effects of expansion and maintain the initial 2 MK temperature. (Note that these arguments do not apply if spicules play only a minor role in the corona.)

[23] Having established that the upflow and downflow have similar temperature, we can interpret equation (11) as the ratio of blue wing to line core intensities under the assumption that all coronal plasma comes from spicules. Table 1 gives the predicted ratios for parameter values $h_c = 5 \times 10^4$ km, $A = 3$, $h_s = 10^4$ km, $v = 100$ km s$^{-1}$, $\delta = 0.1$ (scenario A) or 1 (scenario B), and $n_c = 3 \times 10^9$ cm$^{-3}$ (active region) or $10^9$ cm$^{-3}$ (quiet Sun). The coronal densities are those measured at 1–2 MK by *Dere* [1982], *Doschek et al.* [2007], and *Young et al.* [2009] in active regions, and by *Feldman et al.* [1978], *Laming et al.* [1997], and *Warren and Brooks* [2009] in the quiet Sun. Table 1 also gives predicted values for coronal holes, where we have used a density of $2 \times 10^8$ cm$^{-3}$ and a temperature of 0.8 MK to compute the radiative cooling time [*Del Zanna and Bromage*, 1999; *Doschek et al.*, 1997; *Landi*, 2008]. A more appropriate emission line for coronal holes is Ne VIII (770 Å), which exhibits similar RB asymmetries [*De Pontieu et al.*, 2009].

[24] We see that the predicted ratios are one to two orders of magnitude larger than the observed ratio, $R_{obs} \le 0.05$. For scenario A, which is the better description of most spicules, the predicted ratio exceeds the observed one by at least a factor of 68 in active regions, 22 in the quiet Sun, and 14 in coronal holes. The disagreement is smallest for scenario B in the coronal holes, where the difference is a factor of 5.

[25] There is considerable uncertainty in the predicted ratios. If we allow each of the six parameters in equation (11) to be uncertain by a factor of 2, and if we assume that the errors are uncorrelated, then the combined uncertainty in $R$ is a factor of 3.6. This is much too small to account for the discrepancy with observations. If we allow the parameters to be uncertain by a factor of 4, which seems excessive, then the uncertainty in $R$ is a factor of 9, still too small, with the exception of scenario B in coronal holes and the quiet Sun if we ignore the inequalities.

[26] A potential source of error is the assumption that all of the upflowing plasma is at a temperature where the emission line is sensitive. If this is not the case, so that a substantial portion of the upflow is invisible, then the predicted $R$ must be adjusted downward. We have argued that the upflow cannot be much hotter than 2 MK (1 MK in coronal holes) if the corona is dominated by spicules. *Brooks and Warren* [2012] have obtained an emission measure distribution from the blue wings of spectral lines observed in an active region. It decreases steeply with temperature below a peak at about 1.6 MK. This suggests that our determination of $R$ is not greatly affected by "missing" plasma in the upflow, at least not in active regions.

[27] The velocity of 100 km s$^{-1}$ that we have used in equation (11) is based largely on the Doppler shift of blue wing emission observed near disk center (and also on proper motions observed near the limb). The blueshift in fact represents a combination of the upward expansion of the hot plasma and the initial kick imparted by the ejection. The ejection dominates initially, but the strong deceleration of gravity means that expansion rapidly takes over as the primary cause of the upflow. Recall that the He II emitting material reaches a maximum height of only about $10^4$ km, which is the ballistic height expected for an initial velocity of 70 km s$^{-1}$. An upper limit for the expansion velocity is the sound speed of the heated material, or 230 km s$^{-1}$ at $T = 2$ MK.

[28] We note that the Doppler shift will be smaller than the actual velocity if the strand is inclined to vertical. It is

**Table 2.** Fraction of Hot Coronal Plasma Due to Spicules, $f_s$

|  | Scenario A | Scenario B |
| --- | --- | --- |
| Active Region | <1.4% | <3.4% |
| Quiet Sun | <4.5% | <11% |
| Coronal Hole | <7.4% | <19% |

appropriate to use the Doppler shift value for $v$ in equation (11). $v$ appears in the expression because of the substitution $\tau_u = h_c/v$ for the time it takes the expansion to reach a coronal scale height. If velocity is corrected to account for inclination, the distance traveled must also be corrected, so the effects cancel, and $v$ is the Doppler shift.

[29] Even taking the various uncertainties into account, there remains a large discrepancy between the predicted and observed values of $R$, and this implies that spicules supply only a small fraction of the hot plasma in the corona. It is straightforward to show that the ratio of spicule to non-spicule emission measures is given by $f_s = R_{obs}/R$. Values of $f_s$ based on Table 1 are presented in Table 2. They are all small. Spicules appear to account for less than 1.4% of the coronal plasma in active regions under scenario A. They could account for as much as 19% of the coronal plasma in coronal holes under scenario B, but we must remember that scenario B is relatively uncommon. Note that the predicted $R$ given in equation (11) uses the observed characteristic coronal density, $n_c$. If spicules play only a minor role in the corona, then the density could be much different, and $f_s$ may not be accurate. Of course it cannot approach unity, because then the density must be close to the characteristic density, and we know that this gives an $f_s$ that is very small.

## 2.2. Lower Transition Region (LTR) Emission

[30] As indicated in Figure 1, only the top fraction $\delta$ of the cool spicule gets heated to coronal temperatures in scenario A. Most of the rest gets heated to roughly $10^5$ K and is visible in He II (304 Å) for a time $\tau_{LTR} \approx 300$ s as it falls back to the surface [*De Pontieu et al.*, 2011]. Its time-integrated emission measure is given by

$$EM^*_{LTR} = \tau_{LTR}(1-\delta)n_0^2 h_s, \qquad (16)$$

where we have assumed that its density is the same as the initial density of the hot material, $n_0$. We can compare this with the time-integrated emission measure of the 2 MK downflow in equation (9). Using conservation of mass, equation (10), we obtain the ratio

$$\frac{EM^*_{LTR}}{EM^*_d} > 1.5 \times 10^{-13} \frac{1-\delta}{\delta^2} \frac{h_c}{h_s} An_c \tau_{LTR}. \qquad (17)$$

The actual ratio is likely to significantly exceed this lower limit for several reasons. First, equation (9) is an upper limit because cooling from thermal conduction and enthalpy are ignored. Second, the average density of the spicule may be greater than the density of the tip that gets heated to 2 MK. Third, the spicule material may continue to radiate after it has fallen back to the surface and is no longer visible in He II at the limb (i.e., the lifetime may be longer than 300 s). Finally, equation (17) does not include LTR emission coming from the conventional transition region at the base of the strand that is powered by the thermal conduction and enthalpy fluxes from the cooling coronal plasma. Table 3 gives the predicted ratios for $h_c = 5 \times 10^4$ km, $h_s = 10^4$ km, $A = 3$, $n_c = 3 \times 10^9$ cm$^{-3}$, $\tau_{LTR} = 300$ s, and several different values for $\delta$. The ratios range from 16 to 770.

[31] Let us assume for the moment that all of the plasma that does not reach coronal temperatures (the green material in Figure 1) is heated to the temperature range centered on $T = 0.08$ MK where He II (304 Å) is sensitive. This range has an approximate width $\Delta \log T = 0.3$, as does the temperature sensitivity range of the coronal line used to determine $EM^*_d$. Under this assumption, the ratio of emission measures in equation (17) is approximately equal to the ratio of $T \times DEM(T)$ evaluated at 0.08 MK and 2 MK, where $DEM$ is the differential emission measure. Using the differential emission measure distributions of *Raymond and Foukal* [1982], *Raymond and Doyle* [1981], *Dere and Mason* [1993], and *Landi and Chiuderi Drago* [2008], we obtain observed ratios less than 0.1, as indicated in the third column of Table 3. The ratios are especially small in active regions.

[32] Most of the spicule material is heated above 0.02 MK, since the spicule disappears in Ca II, but some of it may not be heated enough to be visible in He II. To allow for this possibility, we have integrated $DEM(T)$ from the above cited papers between 0.02 and 0.1 MK to obtain an alternative estimate for $EM^*_{LTR}$. For the coronal plasma we use

$$EM^*_d = \ln 10 \, T \, DEM \, \Delta \log T \qquad (18)$$

evaluated at 2 MK with $\Delta \log T = 0.3$. This is approximately equivalent to integrating $DEM(T)$ over a temperature interval corresponding to a temperature decrease of a factor of 2, as would occur during a coronal cooling time. The observed LTR to coronal emission measure ratio obtained in this way is <3 (<1 in active regions), as indicated in the last column of Table 3. There remains a major discrepancy between the predicted and observed ratios. We conclude once again that our starting hypothesis is incorrect, and spicules supply only a minor fraction of the hot plasma in the corona.

[33] The above discussion applies specifically to active regions and the quiet Sun. The conclusion is also valid for coronal holes, though perhaps not as strong. Observed ratios are about a factor of 10 larger than in Table 3 and the predicted ratio is about a factor of 1.5 larger (using 0.8 MK instead of 2 MK). Recall that the predicted ratio is a rather extreme lower limit.

[34] Although spicules seem to account for only a small fraction of the emission from the corona, they may be responsible for most of the emission from the lower transition region. The LTR predicted by standard coronal heating models is much too faint. The $DEM(T)$ in those models decreases with decreasing temperature throughout the transition

**Table 3.** Lower Transition Region to Coronal Emission Measure Ratio

| $\delta$ | Predicted | Observed (He II 304) | Observed ($T \leq 0.1$ MK) |
| --- | --- | --- | --- |
| 0.05 | >770 | <0.1 | <3 |
| 0.1 | >180 | <0.1 | <3 |
| 0.2 | >41 | <0.1 | <3 |
| 0.3 | >16 | <0.1 | <3 |

region, all the way down to the chromosphere. They miss the sharp upturn that is observed below about 0.1 MK. This seems to be true whether the heating is steady or impulsive, though nonequilibrium ionization effects could be important in impulsive heating models, and this has not been fully explored. Small low-lying loops that are everywhere cooler than 0.1 MK have been proposed as an explanation for the excess emission of the LTR [*Antiochos and Noci*, 1986]. Such loops may be present in the mixed polarities of the quiet Sun, but they cannot exist in the large unipolar areas of active regions. Spicules provide a natural explanation for the bright LTR emission.

[35] Spicules may also explain the rather strong redshifts of 10–15 km s$^{-1}$ that are observed in the LTR of the quiet Sun [*Peter and Judge*, 1999] and active regions [*Klimchuk*, 1987]. If the material seen in He II (304 Å) falls at even a fraction of its observed upflow speed, then large redshifts are possible. Parabolic trajectories seen in height-time plots from He II limb observations [*De Pontieu et al.*, 2011] support this conjecture.

### 2.3. High-Frequency Spicules

[36] *De Pontieu et al.* [2011] have proposed a spicule scenario that is significantly different from the two scenarios we have discussed so far. Call it scenario C. A given coronal strand may experience multiple spicule ejections. The frequency of events in scenarios A and B is low in the sense that the coronal plasma from one spicule has time to cool and drain before the next spicule occurs. For typical coronal densities, the cooling time given by equation (8) is roughly 2000 s in active regions and 7000 s in the quiet Sun. Spicules are proposed to occur much more frequently in scenario C. Based on disk observations of H$_\alpha$ rapid blueshift events (RBEs), *De Pontieu et al.* [2011] suggest that the recurrence time at the same location may be as short as 500 s. Using conservative estimates for the density of the ejected hot plasma, they conclude that each event contributes only a few percent of the material in a typical coronal strand. Presumably the mass slowly builds up until coronal densities are reached, at which point continued ejections offset the mass loss from draining. De Pontieu et al. estimate that the flux of kinetic and thermal energy in the spicules is adequate to sustain the radiative and conductive energy losses from the strand once it is fully developed.

[37] We find two significant difficulties with this scenario. The first concerns the evolution of the loop strand. If each spicule contributes a few percent of the eventual mass, then at least 20 spicules are required to reach the final state (more if any draining takes place during the buildup). At a frequency of one per 500 s, approximately 10$^4$ s are required for a fully developed strand to appear. In contrast, observationally distinct loops at 1–2 MK have a total lifetime of only 1000–5000 s, including both the brightening and fading phases [*Klimchuk et al.*, 2010]. Fewer spicules and therefore less time would be needed to fill a loop if the density of the ejected hot plasma were much larger than assumed. The bigger ejections might, however, produce brightness fluctuations that are larger than observed [*De Pontieu et al.*, 2011]. It seems that Scenario C has significant difficulty in simultaneously explaining both the duration and variability of the emission from these loops. This criticism does not apply to soft X-ray loops, which have longer observed lifetimes [*López Fuentes et al.*, 2007], or to the diffuse corona in which individual loops cannot be identified.

[38] Perhaps a greater concern about scenario C is whether it can explain the temperature structure of the corona. The maximum temperature is observed to occur near the strand apex, not near the foot points. Each spicule under scenario C provides hot plasma and energy only to the extreme lower part of the strand, since the newly ejected plasma cannot pass through the pre-existing plasma from the earlier ejections. An upwardly directed thermal conduction flux is therefore required to power the radiation from the overlying material. This implies a temperature inversion, with the maximum temperature occurring near the base. It can be shown that the temperature scale height needed to carry this conduction flux is approximately 1.7 × 10$^4$ km in active regions and 5.1 × 10$^4$ km in the quiet Sun. This can be easily ruled out observationally. Waves may be generated during the spicule ejections, and these might heat the corona at higher elevations and eliminate the need for a temperature inversion, but this would constitute coronal heating [*McIntosh et al.*, 2011].

## 3. Conclusions

[39] The discovery of type II spicules and their association with red-blue asymmetries in EUV spectral lines suggested the interesting possibility that most coronal plasma comes from spicule material that is heated to ≈2 MK as it is ejected. The need for coronal heating—energy deposition in the corona itself—was brought into question. We have presented simple yet physically meaningful arguments that raise serious doubts about this possibility. According to our calculations, spicules supply only a small fraction of the hot plasma that exists in the corona: <2% in active regions, <5% in the quiet Sun, and <8% in coronal holes. The large majority of coronal plasma is likely due to chromospheric evaporation, a fundamentally different process that is a response to coronal heating. Furthermore, even if the coronal plasma were to originate in spicules, coronal heating would still be required to maintain the hot temperature as the material expands.

[40] Our approach has been to assume that all hot plasma is supplied by spicules and then to examine the observational consequences of that assumption. One consequence is that the red-blue asymmetry of spectral lines, which relates to the excess intensity in the blue wing relative to the line core, would be many times larger than observed. The discrepancy is roughly two orders of magnitude in active regions. Another consequence is that the ratio of emission measures in the lower transition region and corona would be much too large, again by about two orders of magnitude. Of course these comparisons must be judged against the simplicity of the model and the assumptions inherent to it. More sophisticated hydrodynamic and MHD simulations must be performed for verification [e.g., *Sterling et al.*, 1993; *Martínez-Sykora et al.*, 2011a; *Judge et al.*, 2012]. Nonetheless, the enormity of the discrepancies suggests that the basic conclusions are correct.

[41] Our conclusions differ considerably from those of *De Pontieu et al.* [2009, 2011] and *McIntosh and De Pontieu* [2009]. They derive energy fluxes that exceed the energy requirements of the quiet Sun and coronal holes and approach those of active regions. We believe these derived fluxes are much too large, as discussed in Appendix B.

[42] We find that the lower transition region would be much too bright if the corona were dominated by spicules, but much too faint if the corona were explained solely on the

basis of standard coronal heating models. This leads us to the following picture. The upper solar atmosphere (transition region and corona) is filled with thin magnetic flux strands that are at or below the resolving capability of our best instruments. Most strands are populated with plasma as a consequence of coronal heating. Interspersed among them are far fewer strands that experience type II spicule ejections. Most of the hot emission that we see comes from strands with coronal heating and no spicules. The faint blue wing component comes from spicules and produces the small red-blue asymmetries. Most of the lower transition region emission also comes from spicules. Their inherent brightness at these temperatures more than compensates for the small filling factor. Some blue wing emission may also be produced by coronal nanoflares, but mostly in lines hotter than 3 MK [*Patsourakos and Klimchuk*, 2006].

[43] There is still much to learn about type II spicules, including how they are created in the first place. Our analysis here has examined only what happens after they are formed. One interesting question is whether spicules are different in mixed magnetic polarity areas of the quiet Sun and unipolar areas of active regions and coronal holes. Magnetic reconnection in mixed polarity regions can involve oppositely directed fields (e.g., between a long flux strand and a small "magnetic carpet" loop), but only "component reconnection" is possible where the field is unipolar. Is this an important difference? Initial reports were that type II spicules are generally similar everywhere on the Sun: active regions, quiet Sun, and coronal holes. If true, this would be further evidence that spicules do not provide most of the coronal plasma, since the properties of the corona are much different in these different regions. On the other hand, the properties of the LTR are also different in these regions, and if the LTR is due primarily to spicules, one might expect the spicules to be different too.

[44] A more specific question concerns the origin of the hot material at the spicule tip. Depending on the sound speed (temperature) of the ambient plasma in the strand, the ejected cold spicule may act like a piston and produce shock-heated plasma at the leading edge. Is this the cause of the observed hot emission? If the ambient plasma is too hot or if the ejection velocity is too slow, a shock will not be produced. This could explain why many spicules do not seem to have significant hot emission (see Appendix B). The substantial length of the hot emission in some spicules could be explained by the shock being at different heights in sub-resolution strands as a consequence of slightly different ejection times or slightly different ejection speeds. New observations from the upcoming Interface Region Imaging Spectrograph (IRIS) mission, combined with detailed modeling, will help us to answer these and other crucial questions concerning this fascinating phenomenon.

## Appendix A: Distinction Between Spicules and Chromospheric Evaporation

[45] We here discuss the differences between spicules and chromospheric evaporation. Some readers may wonder whether they are essentially the same thing. In fact, they are fundamentally different. Only hot material flows into the corona during chromospheric evaporation. It is a response of the lower atmosphere to an increase in the downward thermal conduction flux from the corona, as occurs when there is an increase in the coronal heating rate. (We ignore electron beam driven evaporation for the moment.) Because the transition region is unable to radiate the extra energy, super-hydrostatic pressure gradients develop, and plasma is driven upward. If the heating remains steady, evaporation continues until there is enough density in the corona to produce a balance among radiation, heating, and conduction. This is a static equilibrium. If the heating switches off, as in a nano-flare, the evaporated plasma cools and slowly drains back to the surface [e.g., *Klimchuk*, 2006; *Reale*, 2010].

[46] In order to produce an ejection of cold material, as in a spicule, the thermal conduction energy would need to be deposited at the spicule base, deep in the chromosphere, without heating the material above. This is not how evaporation works. A common misconception is that the enhanced heat flux from the corona passes through the transition region and is deposited in the chromosphere below. In reality, only a very small fraction makes it to the chromosphere. Most of the energy is used up heating the transition region plasma. It is helpful to think of the transition region as a stack of thin layers of different temperature. During evaporation, each layer is heated to progressively higher temperatures until it eventually becomes coronal. New plasma continually "enters" the transition region from the top of the chromosphere.

[47] It is obvious that a coronal heat flux cannot penetrate deep into the chromosphere to eject a spicule. The heat flux is proportional to $T^{5/2}\nabla T$, and both the temperature and temperature gradient are small in the chromosphere. A beam of high-energy particles could potentially deposit a large amount energy deep in the chromosphere, but flare simulations to date have not produced anything that resembles a spicule (G. H. Fisher, personal communication, 2012; J. C. Allred, personal communication, 2012). Instead, they reveal another type of evaporation, termed explosive evaporation, in which all of the upflowing material is heated.

[48] We also point out that the expected behavior of the $\sim$1 MK plasma during coronal nanoflares is the opposite to what is observed in spicules. The 1 MK emission from spicules is observed to rise upward together with the cool jet when viewed near the limb [*De Pontieu et al.*, 2011]. In nanoflares, the 1 MK emission comes from the transition region foot points of super-heated loop strands. The greatly increased pressure in these strands pushes the transition region and chromosphere downward, so the 1 MK emission layer is actually displaced deeper in the atmosphere, even as the plasma itself is rapidly flowing upward from evaporation.

[49] This behavior is demonstrated in two simulations performed with the ARGOS 1D hydrodynamics code [*Antiochos et al.*, 1999]. In both simulations we begin with a semi-circular coronal loop strand of $7.5 \times 10^4$ km halflength that is maintained in a static equilibrium by a uniform $10^{-6}$ erg cm$^{-3}$ s$^{-1}$ volumetric heating rate. Chromospheric sections with many scale heights of $3 \times 10^4$ K plasma are attached at each end. In the first simulation, we set off an impulsive nanoflare corresponding to a rapid increase and decrease in the spatially uniform heating rate. The temporal profile is triangular and lasts a total of 50 s. The peak heating rate is $1.5 \times 10^{-2}$ erg cm$^{-3}$ s$^{-1}$ and the total energy input is $5.625 \times 10^9$ ergs cm$^{-2}$. This simulation is identical to Example 2 in *Klimchuk et al.* [2008] except that heat flux saturation is included in the present case.

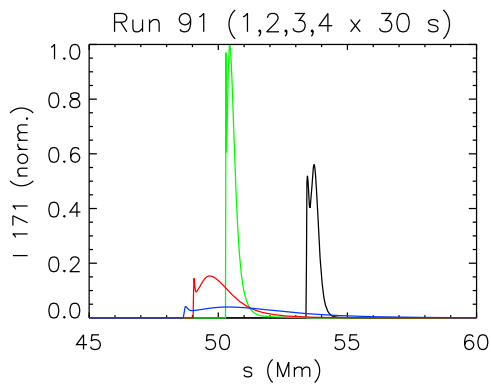

**Figure A1.** Simulated TRACE 171 observation of the foot point region of a loop strand heated by a 50 s nanoflare. Shown are profiles of normalized intensity versus position at times of 30 (black), 60 (green), 90 (red), and 120 s (blue) after the start of the nanoflare. Increasing $s$ corresponds to the upward direction.

[50] Figure A1 shows synthetic observations corresponding to the 171 channel of the Transition Region and Coronal Explorer (TRACE). The bandpass is similar to the 171 channel of SDO and is dominated by lines of Fe IX and X, formed near 1 MK. Intensity is plotted as a function of position near the "left" foot point of the loop (the apex is located at $s = 135$ Mm). The black, green, red, and blue curves correspond to 30, 60, 90, and 120 s after the start of the nanoflare, respectively. We see that the 171 emission feature moves downward with a maximum velocity of about 100 km s$^{-1}$. It lasts for approximately one minute, at which point the intensity and velocity decrease dramatically. A period of slow (<15 km s$^{-1}$) and very faint (normalized intensity <1%) upward motion follows. We note that these velocities are larger than what we would expect on the Sun. The velocity scales with the gravitational scale height in the chromosphere and therefore the chromospheric temperature. The $3 \times 10^4$ K used in our simulations is about a factor of 3 too large.

[51] The second example considers a more gradual nanoflare. The total energy release is the same, but it is now spread out over 500 s. This is the same simulation as Example 1 in *Klimchuk et al.* [2008] and discussed in *Klimchuk* [2006] (the case without saturated heat flux). Figure A2 shows the 171 intensity profiles at 100, 200, 300, and 400 s after the start of the nanoflare. Again, the proper motion of the emission layer is initially downward, this time peaking at about 30 km s$^{-1}$. The subsequent upward displacement is perhaps bright enough to be detected, but it is much slower (<5 km s$^{-1}$) than observed in spicules. It is clear that spicules are not produced by coronal nanoflares. The proper motion of hot emission is wrong, and more importantly, there is no ejection of plasma at chromospheric temperatures.

## Appendix B: Energetics

[52] The arguments we have presented in the main text are entirely independent of energy flux considerations, but it is nonetheless important to examine the energy carried into the corona by spicules. A first point is that most of the energy returns to the surface with the bulk of the material that falls back in either a cold or warm (≤0.1 MK) state. This is not accounted for in the energy estimates of *Moore et al.* [2011]. What matters for the corona is the hot component at the tip of the spicule that continues to rise upward.

[53] A second point is that the initial kinetic energy of the hot material is energetically unimportant. If all of it were converted to thermal energy, the temperature would increase by $\Delta T = 2.4 \times 10^{-9} v^2$. For $v = 100$ km s$^{-1}$, this is only 0.24 MK. Thus, a conversion of the initial kinetic energy imparted by the ejection cannot compensate the severe adiabatic cooling that occurs during the expansion (Section 2.1).

[54] What happens to the thermal energy that is lost during the expansion? Clearly it does not disappear. Some supplies the kinetic energy of the expansion (different from the initial ejection), some does work on pre-existing plasma contained in the strand, some goes into gravitational potential energy, and some is lost to radiation. Suppose the strand were a completely closed system with rigid, thermally insulated caps at the ends. In the absence of gravity and pre-existing material, and ignoring radiation, the expanding plasma would bounce back and forth until the motions are eventually damped out by viscosity. In this idealized case, the kinetic energy would be converted back into thermal energy, and the final temperature would equal the initial temperature at the start of expansion (or slightly higher if there is an ejection). In reality, of course, the system is not closed. Hot material expanding from one foot point will traverse the strand and interact with cold material at the other foot point. Energy will be lost to compression waves propagating into the chromosphere, to new material evaporated into the strand, and to radiation. In the case of coronal holes, energy will be lost to the solar wind outflow. The pre-expansion temperature will never be recovered without coronal heating.

[55] *De Pontieu et al.* [2009, 2011] have presented two estimates of the upward energy flux in the hot material that exceed the energy requirements of the quiet Sun and coronal holes and approach those of active regions. We examine each of these estimates in turn. *De Pontieu et al.* [2009] performed an analysis not unlike our analysis in Section 2.1. A fundamental difference, however, is that they treat the blue wing to line core intensity ratio $R$ as a specified parameter that they set to 5%. We instead treat $R$ as a derived quantity, which we then compare to the observed ratio $R_{obs} = 5\%$. Their energy flux estimate is based on the assumption that all of the observed emission in the line profile, and therefore all of the

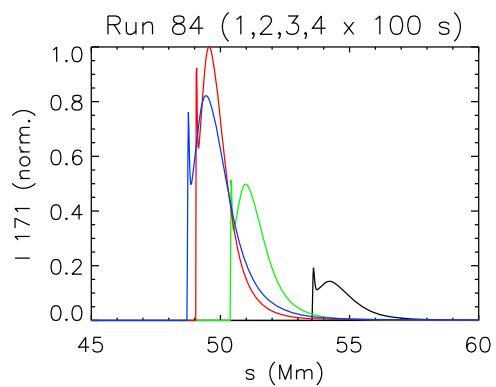

**Figure A2.** As in Figure A1 except for a 500 s nanoflare. Profiles are from 100 (black), 200 (green), 300 (red), and 400 s (blue) after the start of the nanoflare.

plasma in the corona, comes from spicules. If most of the emission in the line core is unrelated to spicules, as we have found, then their energy flux estimate will be incorrect.

[56] *De Pontieu et al.* [2009] also use an alternate form for the equation of mass conservation:

$$n_i v N_i = \frac{n_c h_c A}{\tau_{cool}}, \quad (B1)$$

where $n_i$ is the "density of coronal mass propelled upward and heated in a spicule" and $N_i$ is the "number of spicules that occur at any location over the spicule lifetime." It is clear from the discussion in the paper that $N_i$ is equivalent to the area filling factor of the hot upflow (the fraction of the solar surface undergoing hot upflow at any given time). If we take $n_i$ to be the density of the expanding plasma column when it has vertical extent $h_i$, we can rewrite equation (10) as

$$n_i h_i = n_c h_c A. \quad (B2)$$

Comparing equations (B1) and (B2), we see that $N_i = h_i/(v\tau_{cool})$. Since hot emission is typically observed to protrude a few thousand kilometers above the cold spicule [*De Pontieu et al.*, 2011], we take $h_i = 5000$ km. Together with $v = 100$ km s$^{-1}$ and $\tau_{cool} = 2000$, this gives $N_i = 0.025$. This is up to 80 times smaller than the value arrived at by *De Pontieu et al.* [2009]. The maximum discrepancy occurs when the emission measure distribution of the upflow is reasonably narrow, as found by *Brooks and Warren* [2012].

[57] The expression for $N_i$ derived by *De Pontieu et al.* [2009] is inversely proportional to $R$ (they use the symbol $\alpha$). If $R \gg 0.05$ as we have found, i.e., if only a minority of coronal flux strands are filled by spicules, then their $N_i$ will be much too large, as will the corresponding estimate for the energy flux. We conclude that the estimate must be adjusted downward to approximately $6 \times 10^4$ erg cm$^{-2}$ s$^{-1}$, well below the energy requirements of the quiet Sun and coronal holes. *De Pontieu et al.* [2009] indicate that their large filling factor is compatible with the number density of cold spicules observed by Hinode/SOT. This suggests that only a small fraction of spicules produce hot plasma.

[58] Note that the right side of equation (B1) is equivalent to $n_c v_c N_c$, implying a hot downflow of velocity $v_c = Ah_c/\tau_{cool} = 75$ km s$^{-1}$ for a downflow filling factor $N_c = 1$. Smaller filling factors require even larger velocities. Such downflows are not observed [e.g., *Tripathi et al.*, 2012]. Our conclusion is once again that the hot upflow filling factor $N_i$ derived by *De Pontieu et al.* [2009] and the energy flux based upon it are too large.

[59] *De Pontieu et al.* [2011] take a somewhat different approach to the problem. They introduce a temporal filling factor, $f_t$, which is the fraction of time that hot spicule upflows are present at any given location. They estimate $f_t = 0.2$ based on the observation that H$_\alpha$ rapid blueshift events (RBEs) recur at the same spot roughly every 5–10 min and that coronal events detected by SDO/AIA last up to 2–4 min. We see two potential difficulties with this estimate. First, a majority of spicules/RBEs may not have hot plasma (see above). Second, the length of time that coronal emission is visible during an event may be different from the timescale relevant to the ejection of new material into the corona. Consider a garden hose analogy. If the valve is turned on and off quickly, water will be visible for longer than the time that the valve is in the open position.

[60] While we suggest that *De Pontieu et al.* [2009, 2011] have over-estimated the energy flux of hot spicule material, we emphasize that the observational discrepancies discussed in the main text of the paper are entirely independent of the concerns raised here.

[61] **Acknowledgments.** This work was supported by the NASA Supporting Research and Technology Program. The author thanks many people for useful discussions, but especially Bart De Pontieu. Martin Laming, Scott McIntosh, John Raymond, Karel Schrijver, Alphonse Sterling, Peter Cargill, Steve Bradshaw, and Nicholeen Viall were also particularly helpful.

[62] Philippa Browning thanks Peter Cargill and Petrus C. Martens for their assistance in evaluating this paper.

## References

Antiochos, S. K., and G. Noci (1986), The structure of the static corona and transition region, *Astrophys. J.*, *301*, 440–447.
Antiochos, S. K., P. J. MacNeice, D. S. Spicer, and J. A. Klimchuk (1999), The dynamic formation of prominence condensations, *Astrophys. J.*, *512*, 985–991.
Athay, R. G., and T. E. Holzer (1982), The role of spicules in heating the solar atmosphere, *Astrophys. J.*, *255*, 743–752.
Beckers, J. M. (1972), Solar spicules, *Ann. Rev. Astron. Astrophys.*, *10*, 73–100.
Bradshaw, S. J., and P. J. Cargill (2010), The cooling of coronal plasmas. III. Enthalpy transfer as a mechanism for energy loss, *Astrophys. J.*, *717*, 163–174.
Bradshaw, S. J., J. A. Klimchuk, and J. W. Reep (2012), Diagnosing the time-dependence of active region core heating from the emission measure: I. Low-frequency nanoflares, *Astrophys. J.*, *758*, 53–61, doi:10.1088/0004-637X/758/1/53.
Brooks, D. H., and H. P. Warren (2012), The coronal source of extreme-ultraviolet line profile asymmetries in solar active region outflows, *Astrophys. J. Lett.*, *760*(1), L5, doi:10.1088/2041-8205/760/1/L5.
Del Zanna, G., and B. J. I. Bromage (1999), The elephant's trunk: Spectroscopic diagnostics applied to SOHO/CDS observations of the August 1996 equatorial coronal hole, *J. Geophys. Res.*, *104*(A5), 9753–9766.
De Pontieu, B., et al. (2007), A tale of two spicules: The impact of spicules on the magnetic chromosphere, *Publ. Astron. Soc. Jpn.*, *59*, S655–S662.
De Pontieu, B., S. W. McIntosh, V. H. Hansteen, and C. J. Schrijver (2009), Observing the roots of solar coronal heating—In the chromosphere, *Astrophys. J.*, *701*, L1–L6.
De Pontieu, B., et al. (2011), The origins of hot plasma in the solar corona, *Science*, *331*, 55–58.
Dere, K. P. (1982), Extreme ultraviolet spectra of solar active regions and their analysis, *Solar Phys.*, *77*, 77–93.
Dere, K. P., and H. E. Mason (1993), Nonthermal velocities in the solar transition zone observed with the high-resolution telescope and spectrograph, *Solar Phys.*, *144*, 217–241.
Doschek, G. A. (2012), The dynamics and heating of active region loops, *Astrophys. J.*, *754*, 153–169.
Doschek, G. A., et al. (1997), Electron densities in the solar polar coronal holes from density-sensitive line ratios of Si XIII and S X, *Astrophys. J.*, *482*, L109–L112.
Doschek, G. A., et al. (2007), The temperature and density structure of an active region observed with the Extreme-Ultraviolet Imaging Spectrometer on Hinode, *Publ. Astron. Soc. Jpn.*, *59*, S707–S712.
Feldman, U., G. A. Doschek, J. T. Mariska, A. K. Bhatia, and H. E. Mason (1978), Electron densities in the solar corona from density-sensitive line ratios in the N I isoelectronic sequence, *Astrophys. J.*, *226*, 674–678.
Hansteen, V. H., H. Hara, B. De Pontieu, and M. Carlsson (2010), On redshifts and blueshifts in the transition region and corona, *Astrophys. J.*, *718*, 1070–1078.
Hara, H., et al. (2008), Coronal plasma motions near footpoints of active region loops revealed from spectroscopic observations with Hinode EIS, *Astrophys. J.*, *678*, L67–L71.
Judge, P. G., B. De Pontieu, S. W. McIntosh, and K. Olluri (2012), The connection of type II spicules to the corona, *Astrophys. J.*, *746*, 158–166.
Klimchuk, J. A. (1987), On the large-scale dynamics and magnetic structure of solar active regions, *Astrophys. J.*, *323*, 368–379.
Klimchuk, J. A. (2006), On solving the coronal heating problem, *Solar Phys.*, *234*, 41–77.

Klimchuk, J. A. (2011), Are spicules the primary source of hot coronal plasma?, *Bull. Am. Astron. Soc.*, *43*, Abstract 18.01.

Klimchuk, J. A., Patsourakos, S., and Cargill, P. J. (2008), Highly efficient modeling of dynamic coronal loops, *Astrophys. J.*, *682*, 1351–1362.

Klimchuk, J. A., J. T. Karpen, and S. K. Antiochos (2010), Can thermal nonequilibrium explain coronal loops?, *Astrophys. J.*, *714*, 1239–1248.

Laming, J. M., et al. (1997), Electron density diagnostics for the solar upper atmosphere from spectra obtained by SUMER/SOHO, *Astrophys. J.*, *485*, 911–919.

Landi, E. (2008), The off-disk thermal structure of a polar coronal hole, *Astrophys. J.*, *685*, 1270–1276.

Landi, E., and F. Chiuderi Drago (2008), The quiet-Sun differential emission measure from radio and UV measurements, *Astrophys. J.*, *675*, 1629–1636.

López Fuentes, M. C., J. A. Klimchuk, and C. H. Mandrini (2007), The temporal evolution of coronal loops observed by GOES SXI, *Astrophys. J.*, *657*, 1127–1136.

Madjarska, M. S., Vanninathan, K., and Doyle, J. G. (2011), Can coronal hole spicules reach coronal temperatures?, *Astron. Astrophys.*, *532*, L1–L4.

Martínez-Sykora, J., B. De Pontieu, V. H. Hansteen, and S. W. McIntosh (2011a), What do spectral line profile asymmetries tell us about the solar atmosphere?, *Astrophys. J.*, *732*, 84–109.

Martínez-Sykora, J., V. Hansteen, and F. Moreno-Insertis (2011b), On the origin of the type II spicules: Dynamic three-dimensional MHD simulations, *Astrophys. J.*, *736*, 9–20.

McIntosh, S. W., and B. De Pontieu (2009), High-speed transition region and coronal upflows in the quiet Sun, *Astrophys. J.*, *707*, 524–538.

McIntosh, S. W., B. De Pontieu, M. Carlsson, V. Hansteen, P. Boerner, and M. Goossens (2011), Alfvenic waves with sufficient energy to power the quiet solar corona and fast solar wind, *Nature*, *475*, 477–480.

Moore, R. L., A. C. Sterling, J. W. Cirtain, and D. A. Falconer (2011), Solar X-ray jets, type-II spicules, granule-size bipoles, and the genesis of the heliosphere, *Astrophys. J. Lett.*, *731*, L18–L22.

Patsourakos, S., and J. A. Klimchuk (2006), Nonthermal spectral line broadening and the nanoflare model, *Astrophys. J.*, *647*, 1452–1465.

Patsourakos, S., and J. A. Klimchuk (2009), Spectroscopic observations of hot lines constraining coronal heating in solar active regions, *Astrophys. J.*, *696*, 760–765.

Peter, H., and P. G. Judge (1999), On the Doppler shifts of solar ultraviolet emission lines, *Astrophys. J.*, *522*, 1148–1166.

Raymond, J. C., and J. G. Doyle (1981), The energy balance in coronal holes and average quiet-Sun regions, *Astrophys. J.*, *247*, 686–691.

Raymond, J. C., and P. Foukal (1982), The thermal structure of coronal loops and implications for physical models of coronae, *Astrophys. J.*, *253*, 323–329.

Reale, F. (2010), Coronal loops: Observations and modeling of confined plasma, *Living Rev. Solar Phys.*, *7*, 5–78.

Reale, F., P. Testa, J. A. Klimchuk, and S. Parenti (2009), Evidence of widespread hot plasma in nonflaring coronal active region from Hinode/X-Ray Telescope, *Astrophys. J.*, *698*, 756–765.

Rouppe van der Voort, L., J. Leenaarts, B. De Pontieu, M. Carlsson, and G. Vissers (2009), On-disk counterparts of type II spicules in the Ca II 854.2 nm and $H_\alpha$ lines, *Astrophys. J.*, *705*, 272–284.

Schmelz, J. T., and S. Pathak (2012), The cold shoulder: Emission measure distributions of active region cores, *Astrophys. J.*, *756*(2), 126, doi:10.1088/0004-637X/756/2/126.

Sterling, A. C. (2000), Solar spicules: A review of recent models and targets for future observations, *Solar Phys.*, *196*, 79–111.

Sterling, A. C., K. Shibata, and J. T. Mariska (1993), Solar chromospheric and transition region response to energy deposition in the middle and upper chromosphere, *Astrophys. J.*, *407*, 778–789.

Sterling, A. C., L. K. Harra, and R. L. Moore (2010), Fibrillar chromospheric spicule-like counterparts to an extreme-ultraviolet and soft X-ray blowout coronal jet, *Astrophys. J.*, *722*, 1644–1653.

Sturrock, P. A., W. W. Dixon, J. A. Klimchuk, and S. K. Antiochos (1990), Episodic coronal heating, *Astrophys. J.*, *356*, L31–L34.

Tian, H., et al. (2011), Two components of the solar coronal emission revealed by extreme-ultraviolet spectroscopic observations, *Astrophys. J.*, *738*, 18–27.

Tripathi, D., J. A. Klimchuk, and H. E. Mason (2011), Emission measure distribution and heating of two active region cores, *Astrophys. J.*, *740*, 111–120.

Tripathi, D., H. E. Mason, and J. A. Klimchuk (2012), Active region moss: Doppler shifts from Hinode/Extreme-ultraviolet Imaging Spectrometer observations, *Astrophys. J.*, *753*, 37–45, doi:10.1088/0004-637X/753/1/37.

Vanninathan, K., M. S. Madjarska, E. Scullion, and J. G. Doyle (2012), Off-limb (spicule) DEM distribution from SoHO/SUMER observations, *Solar Phys.*, *280*(2), 425–434, doi:10.1007/s11207-012-9986-8.

Warren, H. P., and D. H. Brooks (2009), The temperature and density structure of the quiet Sun. I. Observations of the quiet Sun with the EUV Imaging Spectrometer on Hinode, *Astrophys. J.*, *700*, 762–773.

Warren, H. P., A. R. Winebarger, and D. H. Brooks (2012), A systematic study of high temperature emission in solar active regions, *Astrophys. J.*, in press.

Winebarger, A. R., J. T. Schmelz, H. P. Warren, S. H. Saar, and V. L. Kashyap (2011), Using a differential emission measure and density measurements in an active region core to test a steady heating model, *Astrophys. J.*, *740*, 2, doi:10.1088/0004-637X/740/1/2.

Young, P. R., T. Watanabe, H. Hara, and J. T. Mariska (2009), High-precision density measurements in the solar corona I. Analysis methods and results for Fe XII and Fe XIII, *Astron. Astrophys.*, *495*, 587–606.

Zhang, Y. Z., K. Shibata, J. X. Wang, X. J. Mao, T. Matsumoto, Y. Liu, and J. T. Su (2012), Revision of solar spicule classification, *Astrophys. J.*, *750*, 16–24.